\documentclass[aps,prc,showpacs,superscriptaddress,byrevtex,
floatfix]{revtex4-1}

\usepackage{graphicx}
\usepackage{color}
\usepackage{amsmath}
\usepackage{braket}
\usepackage{hyperref}
\usepackage{bm}

\hypersetup{bookmarksnumbered,%
  colorlinks,%
  linkcolor=blue,%
  citecolor=blue,%
  plainpages=false,%
  pdfstartview=FitH}

\newcommand{\nc}{\newcommand}           

\nc{\figwidth}{0.5}                     
\nc{\densfigwidth}{0.45}           
\nc{\corfigwidth}{0.28}            

\usepackage{color}
\usepackage{ulem}
\nc {\ir} [1] {\textcolor{red}{#1}}
\nc {\il} [1] {\textcolor[RGB]{0,0,200}{#1}} 
\nc {\ig} [1] {\textcolor[RGB]{0,100,0}{#1}}
\nc {\ib} [1] {\textcolor[RGB]{0,150,255}{(#1)}}
\nc {\im} [1] {\textcolor{magenta}{#1}}

\nc {\irs} [1] {\textcolor{red}{\sout{#1}}}
\nc {\e}{\texorpdfstring{\MakeLowercase{e}}{e}}

\usepackage{soul}
\nc{\mydraft}  {\setlength{\topmargin}{-1.5cm}}
\mydraft   
\begin{document}

\title{Direct probing of the cluster structure in ${}^{12}$Be via
  $\alpha$-knockout reaction }

\author{Mengjiao Lyu} \email[]{mengjiao@rcnp.osaka-u.ac.jp}
\affiliation{Research Center for Nuclear Physics (RCNP), Osaka University,
  Ibaraki 567-0047, Japan}

\author{Kazuki Yoshida} \email[]{yoshidak@rcnp.osaka-u.ac.jp}
\affiliation{Advanced Science Research Center, Japan Atomic Energy Agency,
  Tokai, Ibaraki 319-1195, Japan}

\author{Yoshiko Kanada-En'yo} \email[]{yenyo@ruby.scphys.kyoto-u.ac.jp}
\affiliation{Department of Physics, Kyoto University, 
  Kyoto 606-8502, Japan}

\author{Kazuyuki Ogata} \email[]{kazuyuki@rcnp.osaka-u.ac.jp}
\affiliation{Research Center for Nuclear Physics (RCNP), Osaka University,
  Ibaraki 567-0047, Japan}
\affiliation{Department of Physics, Osaka City University, Osaka 558-8585,
  Japan}
\affiliation{Nambu Yoichiro Institute of Theoretical and Experimental Physics (NITEP),\\ Osaka City University, Osaka 558-8585, Japan}

\begin{abstract}
\begin{description}
\item[Background] Recent theoretical and experimental researches using
  proton-induced $\alpha$-knockout reactions provide direct manifestation of
  $\alpha$-cluster formation in nuclei. In recent and future experiments,
  $\alpha$-knockout data are available for neutron-rich beryllium isotopes. In
  $^{12}$Be , rich phenomena are induced by the formation of $\alpha$-clusters
  surrounded by neutrons, for instance, breaking of the neutron magic number
  $N=8$.

\item[Purpose] Our objective is to provide direct probing of the
  $\alpha$-cluster formation in the $^{12}$Be target through associating the
  structure information obtained by a microscopic theory with the experimental
  observables of $\alpha$-knockout reactions.

\item[Method] We formulate a new wave function of the
  Tohsaki-Horiuchi-Schuck-R{\"o}pke (THSR) type for the structure calculation
  of ${}^{12}$Be nucleus and integrate it with the distorted wave impulse
  approximation framework for the $\alpha$-knockout reaction calculation of
  $^{12}$Be$(p,p\alpha)^{8}$He.

\item[Results] We reproduce the low-lying spectrum of the $^{12}$Be nucleus
  using the THSR wave function and discuss the cluster structure of the ground
  state. Based on the microscopic wave function, the optical potentials and
  $\alpha$-cluster wave function are determined and utilized in the
  calculation of ${}^{12}$Be($p,p\alpha$)${}^{8}$He reaction at 250 MeV. The
  possibility of probing the clustering state of $^{12}$Be through this
  reaction is demonstrated by analysis of the triple differential cross
  sections that are sensitively dependent on the $\alpha$-cluster amplitude at
  the nuclear surface. 

\item[Conclusions] This study provides a feasible approach to validate
  directly the theoretical predictions of clustering features in the $^{12}$Be
  nucleus through the $\alpha$-knockout reaction.
\end{description}
\end{abstract}

\begin{flushright}
\normalsize
~~~~
NITEP 8\\
February, 2019\\
\end{flushright}

\vspace{15pt}

\maketitle

\section{Introduction}
In atomic nuclei, the clusters emerge as a result of the competition between
the short-range repulsion and the medium-range attraction induced by the Pauli
blocking effect and the properties of nuclear forces \cite{freer18}.
Especially, the $\alpha$-clustering effect is prevalent in nuclear clustering
states because of the spin-isospin saturation in the nucleon-nucleon
interaction. For the description of $\alpha$-clustering states, various
structural theories have been formulated, as introduced in
Refs.~\cite{freer18, enyo01, itagaki01, oertzen06, enyo12, horiuchi12, ito14,
ren18} and references therein. 

In the Hoyle state of $^{12}$C, the $\alpha$-cluster formation has been well
established and the description of clustering state has been treated elegantly
in nuclear theory \cite{tohsaki01}. However, in neutron-rich nuclei, the
description of the $\alpha$-clustering states is more challenging because of
the existence of valence neutrons surrounding $\alpha$-clusters, as shown in
the previous studies of Beryllium isotopes \cite{Oka77, Sey81, Des89, Oer96,
Ara96, Dot97, Kan99, Oga00, itagaki00, Des02, ito04, ito08, ito12, enyo03,
kobayashi12, enyo16}. Especially, in $^{10}$Be and $^{12}$Be isotopes, the
nuclear molecular orbit (MO) configuration and the ion-like binary cluster
configuration could coexist in clustering states, as predicted by theoretical
studies using the generalized two-center cluster model (GTCM) \cite{ito04,
ito08, ito12} and antisymmetrized molecular dynamics (AMD) \cite{enyo03,
kobayashi12, enyo16}. In the $^{12}$Be nucleus, the breaking of the neutron
magic number $N$=8 also occurs as a consequence of $\alpha$-cluster formation
\cite{ito08,ito12,enyo03}.

In previous decades, the $\alpha$-clustering states in stable nuclei have been
investigated through the proton-induced $\alpha$-knockout reactions
\cite{Roos77, Nadasen80, Carey84, Wang85, Nadasen89, Mabiala09, yoshida16,
wakasa17, yoshida16, lyu18, yoshida18}. The significant advantage in these
studies is that the physical observables  are directly connected to the
$\alpha$-clusters \cite{yoshida16, lyu18, yoshida18}, and the reaction
mechanism is clean as compared to the other direct reactions where
$\alpha$-clusters are involved, such as the $\alpha$-transfer reactions
\cite{Becchetti1978, Anantaraman1979, Tanabe1981, fukui16}. The theoretical
description of $\alpha$-knockout reactions has been formulated using the
distorted-wave impulse approximation (DWIA) framework \cite{Roos77, wakasa17,
yoshida16, yoshida18}, and in recent works \cite{yoshida16,yoshida18} the
peripheral property of the $(p,p\alpha)$ reactions has been demonstrated. This
is essential for probing $\alpha$-clusters, which are most probably formed in
the surface region of nuclei. Recently, there are emerging ($p,p\alpha$)
reactions in inverse kinematics for light unstable nuclei including the
neutron-rich Be isotopes that is conducted or planned in the Radioactive
Isotope Beam Factory (RIBF) \cite{yang17}. These experiments provide ideal
opportunities to investigate the clustering states of the neutron-rich Be
isotopes by comparing theoretical predictions of ($p,p\alpha$) reaction
observables and the corresponding experimental results.

In our previous work, we have investigated the $^{10}$Be($p,p\alpha$)$^{6}$He
reaction at 250 MeV by integrating the microscopic description of the
$^{10}$Be target and the $^{6}$He residual nuclei into the DWIA framework for
$\alpha$-knockout reaction, and predicted the triple differential cross
sections (TDX) as a useful observable probing the $\alpha$ clustering in the
$^{10}$Be nucleus \cite{lyu18}. For the structure calculation of the ground
state of $^{10}$Be and $^{6}$He, Tohsaki-Horiuchi-Schuck-R\"opke (THSR) wave
functions have been formulated based on the previous studies \cite{tohsaki01,
Fun15, Fun15a, Fun16, zhou13, zhou14, zhou16, lyu15, lyu16, zhao18}. 

In this work, we further extend the THSR wave function for the $^{12}$Be
nucleus. The theoretical description of the clustering features of the
$^{12}$Be nucleus is not so simple as those of the $^{10}$Be nucleus because
of the coexistence of binary cluster and MO configurations. It is essential to
take into account these different cluster configurations for the description
of the ground state of $^{12}$Be nucleus, in particular, the phenomena of the
$N=8$ magic number breaking. In this study, we show that the TDX can be the
direct probe for such exotic clustering features in the $^{12}$Be nucleus. In
addition, this work provides the new formulation of the THSR wave function for
neutron-rich nucleus $^{12}$Be, which could be utilized in the further studies
of other nuclei near the neutron drip line.

This article is organized as follows. In Section II, we recapitulate the DWIA
framework for the $\alpha$-knockout reaction and the calculation of triple
differential cross sections (TDX). In Section III, we formulate the THSR wave
functions for the $^{12}$Be target and the $^{8}$He residual, and the
extraction of the $\alpha$-cluster wave function. Some details of the
formulation are given in the appendix. In Section IV, we discuss the numerical
results for the nuclear structure of $^{12}$Be and the predictions of
$^{12}$Be($p,p\alpha$)$^{8}$He reaction observables (TDXs). Last Section V
contains the conclusion.

\section{DWIA framework for the ${}^{12}$B\e($p$,$p\alpha$)${}^{8}$H{\e}
         reaction}
\label{secdwia}

\begin{figure}[htbp]
  \begin{center}
    \includegraphics[width=\corfigwidth\textwidth]{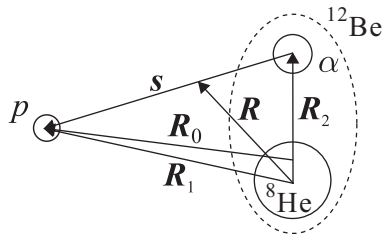}
    \caption{Coordinates of the ${}^{12}$Be($p$,$p\alpha$)${}^{8}$He reaction.
      }
    \label{fig:coordinate}
  \end{center}
\end{figure}

We adopt the same DWIA framework as in Refs.~\cite{yoshida16,yoshida18} for
the ($p,p\alpha$) reaction. In this section, we introduce briefly the DWIA
framework for the $^{12}$Be($p$,$p\alpha$)$^{8}$He reaction. The coordinates
for the description of the $\alpha$-knockout reaction are presented in
Fig.~\ref{fig:coordinate}. Here, the normal kinematics is adopted for
simplicity. The transition amplitude for the ($p$,$p\alpha$) reaction is given
by
\begin{align}
   & T_{{\bm K}_0{\bm K}_1{\bm K}_2}
  = \nonumber                        \\
   & \left<
  \chi_{1,{\bm K}_1}^{(-)}({\bm R}_1)\chi_{2,{\bm K}_2}^{(-)} ({\bm R}_2)
  \left| t_{p\alpha}({\bm s}) \right|
  \chi_{0,{\bm K}_0}^{(+)}({\bm R}_0) \varphi_{\alpha}({\bm R}_2)
  \right>,
  \label{kotmtx}
\end{align}
where the $\chi$ with subscripts 0, 1, and 2 denote the distorted wave
functions for the incident proton $p$, the outgoing $p$, and the outgoing
$\alpha$, respectively. The superscripts $(+)$ and $(-)$ indicate the outgoing
and incoming boundary conditions adopted for $\chi$, respectively. The
$\varphi_{\alpha}$ is the $\alpha$-cluster wave function inside the target
nucleus $^{12}$Be, where only the $0^{+}$ channel is included. For each
particle $i=0,1,2$, the momentum (wave number) and its solid angle in the
center-of-mass frame are denoted by ${\bm K}_{i}$ and $\Omega_{i}$,
respectively, and the corresponding quantities measured in the laboratory
frame are denoted by additional superscript $\rm{L}$. We follow the
theoretical approach in Ref.~\cite{yoshida16} for the numerical calculation of
the triple differential cross section (TDX) of the
${}^{12}$Be($p,p\alpha$)${}^{8}$He reaction: 
\begin{align}\label{eq:tdx}
  \frac{d^3 \sigma}{dE_1^{\rm L} d\Omega_1^{\rm L} d\Omega_2^{\rm L}}
  =&F_{\rm kin}C_0
   \frac{d\sigma_{p\alpha}}{d\Omega_{p\alpha}}
  \left| \bar{T}_{{\bm K}_0{\bm K}_1{\bm K}_2} \right| ^2, \\
  \bar{T}_{{\bm K}_0{\bm K}_1{\bm K}_2}
  =&\int d\bm{R}\,
   \chi_{1,{\bm K}_1}^{(-)}({\bm R})
   \chi_{2,{\bm K}_2}^{(-)} ({\bm R})\nonumber \\ 
  &\quad\times
   \chi_{0,{\bm K}_0}^{(+)}({\bm R})\varphi_{\alpha}(\bm{R})
   e^{-i\bm{K}_0\cdot\bm{R}\,A_\alpha / A},
\end{align}
where $A_\alpha=4$ and $A=12$. $F_{\rm kin}$ and $C_0$ are kinematical
factors, and ${d\sigma_{p\alpha}}/{d\Omega_{p\alpha}}$ is the $p$-$\alpha$
differential cross section at the energy and the scattering angle deduced from
the ($p$,$p\alpha$) kinematics. $\bar{T}_{{\bm K}_0{\bm K}_1{\bm K}_2}$ is a
reduced transition amplitude obtained by making the factorization
approximation to Eq.~(\ref{kotmtx}); details can be found in
Ref.~\cite{yoshida16}.

In this calculation, the optical potentials for the $p$-$^{12}$Be,
$p$-$^{8}$He, and $\alpha$-$^{8}$He systems and the transition interaction
$t_{p\alpha}$ between $p$ and $\alpha$ are determined by the folding model
using the Melbourne $G$-matrix interaction \cite{Amo00}. The density
distributions of the target and residual nuclei are extracted from the THSR
wave function, which is formulated in Section \ref{sec:thsr}, and the
phenomenological density distribution is adopted for the $\alpha$ cluster as
introduced in Ref.~\cite{yoshida16}. The spin-orbit part of each optical
potential was disregarded. The distorted wave functions $\chi_{i}$ ($i=1,2,3$)
are obtained by solving the corresponding Schr{\"o}dinger equations using the
optical potentials mentioned above. The $\alpha$-cluster wave function
$\varphi_{\alpha}$ is extracted from the THSR wave function of ${}^{12}$Be by
approximating the reduced width amplitude (RWA), as introduced in Section
\ref{subsec:rwa}.

\section{THSR wave function for the target and residual nuclei}
\label{sec:thsr}

\begin{figure}[hb]
  \centering
  \includegraphics[width=\densfigwidth\textwidth]{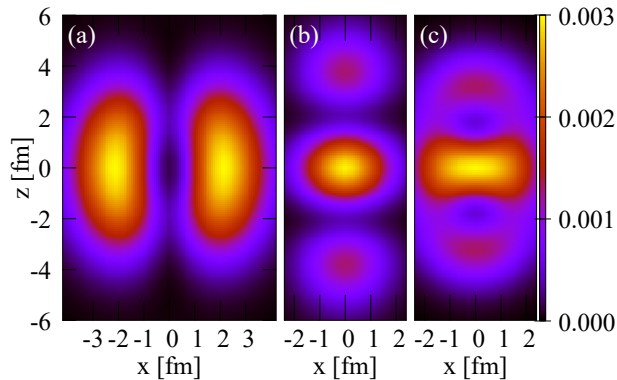}
  \caption{The panels (a) and (b) show the distributions of valance neutrons
    of $^{12}$Be occupying $\pi$- and $\sigma$- molecular orbits,
    respectively, as defined in Eqs.~(\ref{eq:pi-orbit2}) and
    (\ref{eq:sigma-integrated}). The panel (c) shows the density distribution
    of valance neutron occupying the vertical $p$-orbit states of the $^{8}$He
    cluster in the $\alpha$+$^{8}$He configuration of $^{12}$Be, as defined in
    Eq.~(\ref{eq:12Be-2cluster}) and (\ref{eq:p-orbit-vertical}). Parameters
    for each configuration are listed in Table \ref{table:para} in the
    Appendix.}
  \label{fig:dens}
\end{figure}

We formulate the THSR wave functions for the target and residual nuclei by
extending the microscopic models developed in previous works \cite{lyu15,
lyu16, zhao18}. For the target nucleus $^{12}$Be, we consider three kinds of
cluster configurations suggested in the theoretical works \cite{enyo03,ito08}
as basis states in the THSR framework of nonlocalized cluster motion
\cite{zhou13,zhou14}. One is the binary cluster configuration
\begin{equation}
  \alpha+{}^{8}\textrm{He},
\end{equation}
and the other two are the MO configurations
\begin{equation}
  \begin{aligned}
    \pi\text{-orbit:}\quad    & 2\alpha+2n(\pi)+2n(\pi*),   \\
    \sigma\text{-orbit:}\quad & 2\alpha+2n(\pi)+2n(\sigma).
  \end{aligned}
\end{equation}
It has been suggested in Refs.~\cite{ito08} that the binary cluster
configuration $\alpha$+$^{8}$He dominates in $^{12}$Be when the two
$\alpha$-clusters are well separated, while the MO configurations contribute
when the $\alpha$-$\alpha$ distance is smaller than 6 fm. In what follows, we
refer these three configurations as ``$\alpha$+$^{8}$He", ``$\pi$-orbit", and
``$\sigma$-orbit", respectively.

\subsection{The $\alpha$+$^{8}$He configuration of $^{12}$Be}
In this work, we describe the $\alpha$+$^{8}$He configuration of $^{12}$Be in
the THSR framework, as
\begin{equation}
  \begin{aligned}
    \label{eq:12Be-2cluster}
     & \ket{\Phi_{\alpha+{}^{8}\textrm{He}}}= \\
     & \quad\int d\bm{R}\,
    \mathcal{G}(\bm{R},\bm{\beta}_{\alpha})
    \ket{\mathcal{A}\{\Phi_{\alpha}(-\bm{R})\Phi({}^8\textrm{He},\bm{R})\}},
  \end{aligned}
\end{equation}
where the formulation of the $^{8}$He cluster wave function
$\Phi({}^8\textrm{He},\bm{R})$ is explained in the Appendix. Similar
formulation for the $\alpha$+$^{16}$O configuration has been proved to be very
efficient in describing the ground state of $^{20}$Ne in
Refs.~\cite{zhou13,zhou14}. We note that in the THSR framework, each basis
wave function expresses not localized $\alpha$ and $^{8}$He clusters but
non-localized clusters with almost free motion, which is different from the
basis states used in other models \cite{itagaki00,ito08}.

\subsection{The molecular orbit configurations of $^{12}$Be}
The $\pi$-orbit configuration of $^{12}$Be is written as
\begin{equation}
  \begin{aligned}
    \label{eq:12Be-pi}
     & \ket{\Phi_{\pi\text{-orbit}}}= \\
     & \quad\int d\bm{R}
    \mathcal{G}(\bm{R},\bm{\beta}_{\alpha})
    \ket{\mathcal{A}\{\Phi_{\alpha}(-\bm{R})\Phi_{\alpha}(\bm{R})
      \phi_{10}^{\pi}\phi_{10}^{\pi}
      \phi_{11}^{\pi*}\phi_{12}^{\pi*}
      \}},
  \end{aligned}
\end{equation}
where the function $\mathcal{G}$ is the deformed Gaussian for describing the
nonlocalized motion of two $\alpha$-clusters within the $^{12}$Be nucleus,
which is defined by
\begin{equation}\label{eq:Gauss}
  \mathcal{G}(\bm{R},\bm{\beta})=
  \exp \left(-\frac{R_{x}^{2}
    +R_{y}^{2}}{\beta_{xy}^{2}}
  -\frac{R_{z}^{2}}{\beta_{z}^{2}}
  \right).
\end{equation}
The four valance neutrons occupying the $\pi$-orbits are described by the
$\phi_{9,10}^{\pi}$ and $\phi_{11,12}^{\pi*}$ states, which correspond to the
parallel and antiparallel spin-orbit couplings, respectively. The formulation
of single nucleon states $\phi_{9\mbox{--} 12}$ are explained in the Appendix.
In Fig.~\ref{fig:dens} (a), the density distribution is presented for
$\phi_{11,12}$, in which the typical structure of the $\pi$-orbit
configuration is clearly demonstrated. This $\pi$-orbit configuration goes to
the $p$-shell closed configuration in the compact limit of the
$\alpha$-cluster and valence neutron motions.

The $\sigma$-orbit configuration of $^{12}$Be is formulated as
\begin{equation}
  \begin{aligned}
    \label{eq:12Be-sigma}
     & \ket{\Phi_{\sigma\text{-orbit}}}= \\
     & \quad\int d\bm{R}\,
    \mathcal{G}(\bm{R},\bm{\beta}_{\alpha})
    \ket{\mathcal{A}\{\Phi_{\alpha}(-\bm{R})\Phi_{\alpha}(\bm{R})
      \phi_{9}^{\pi}\phi_{10}^{\pi}
      \phi_{11}^{\sigma}\phi_{12}^{\sigma}\}},
  \end{aligned}
\end{equation}
where $\phi_{9,10}^{\pi}$ are the same $\pi$-states as in
Eq.~(\ref{eq:12Be-pi}) and $\phi_{11,12}^{\sigma}$ are states for valance
neutrons occupying the $\sigma$-orbits. The formulation of $\phi_{11, 12}$ are
explained in the Appendix. We also show the density distributions of
$\phi_{11,12}^{\sigma}$ in Fig.~\ref{fig:dens} (b), where the typical nodal
structure in the $\sigma$-orbit is reproduced. As seen in Fig.~\ref{fig:dens}
(c), the $\alpha+{}^{8}\textrm{He}$ configuration also has a similar nodal
structure of the valence neutrons along the $x=0$ axis as a result of the
antisymmetrization effect between neutrons in the $\alpha$ and $^8$He
clusters. In fact, the $\sigma$-orbit configuration is redundant in the
present framework as it is already included in model space when the THSR bases
of the $\alpha$+$^{8}$He configuration are superposed, as discussed in the
Subsection \ref{subsec:energy}.

\subsection{Total wave function and $\alpha$-cluster wave function of target
            nucleus}
\label{subsec:rwa}

The total wave function of $^{12}$Be is obtained by superposing the basis
states in the three configurations formulated in
Eqs.~(\ref{eq:12Be-2cluster}), (\ref{eq:12Be-pi}), and (\ref{eq:12Be-sigma}).
For each configuration, we formulate basis wave functions with different
$\beta_{\alpha,z}$ parameters that manipulate the motion of clusters, and set
other parameters to be the variationally optimized values for each
configuration. All the parameters in the THSR bases are listed in Table
\ref{table:para} in the Appendix. The translational and rotational projections
are performed for the bases to restore corresponding symmetry, as introduced
in Ref.~\cite{lyu15}. With these bases, the total wave function of $^{12}$Be
can be written as 
\begin{equation}\label{eq:12be-total}
  \ket{\Psi^{J}(^{12}\text{Be})}=\sum_{m,j} c_{m,j}
  \hat{P}^{J}_{00}\hat{P}_{\text{c.o.m}}\Phi_{m}(\beta_{\alpha,z;j}),
\end{equation}
where $\Phi_{m}$ labeled by $m$ for cluster configurations are the THSR bases
for $^{12}$Be and $j$ denotes the choice of the parameter $\beta_{\alpha,z}$.
The operators $\hat{P}^{0}_{00}$ and $\hat{P}_{\text{c.o.m}}$ denote angular
momentum projection \cite{schuck} and the projection for center-of-mass motion
\cite{Oka77}, respectively. The $c_{i,j}$ are superposition coefficients to be
obtained by diagonalizing the Hamiltonian matrix.

To extract the $\alpha$-cluster amplitude in the surface region, we
approximate RWA $y(a)$ of the $\alpha$-cluster by the overlap of the total
wave function of $^{12}$Be in Eq.~(\ref{eq:12be-total}) with the
$\alpha$+$^{8}$He cluster wave function as
\begin{equation}
  \label{eq:approx}
  \begin{aligned}
    \left| ay(a) \right| \approx & a y^\textrm{app}(a) \\
    \equiv                       & N_{c}
    \left|\braket{ \Phi({}^{12}\textrm{Be})| \Phi^{{
            (0+)}}_{\textrm{BB}}({}^{8}\textrm{He},\alpha,S=a) }\right|,
  \end{aligned}
\end{equation}
where 
\begin{equation}
  N_{c}=\frac{1}{\sqrt{2}}
  \left(\frac{8\times4}{12\pi b^2}\right)^{1/4}
\end{equation}
with $\Phi_{\textrm{BB}}({}^{8}\textrm{He},\alpha,S)$ being a Brink-Bloch-type
wave function~\cite{Bri66} for the $\alpha$+$^{8}$He two-body system separated
with the relative distance $S$:
\begin{equation}
  \begin{aligned}
     & \ket{\Phi^{(0+)}_{\textrm{BB}}({}^{8}\textrm{He},\alpha,S)} \\
     & \quad=
    \hat{P}^0_{00}\ket{\phi(\alpha,\frac{8}{12}S\vec{e}_z)
      \Phi({}^{8} {\textrm{He}(0^+)},-\frac{4}{12}S\vec{e}_z)
    }.
  \end{aligned}
\end{equation}
Here $\Phi({}^{8} {\textrm{He}(0^+)},-\frac{4}{12}S\vec{e}_z)$ is the wave
function of the residual nucleus $^{8}$He projected onto the $0^{+}$ state,
which is located at $-({4}/{12})S\vec{e}_z$. The $y^{\text{app}}(a)$ is found
to be a good approximation of the exact $y(a)$ in the surface region
\cite{enyo14} and applicable to the present case because the observables in
knockout reactions are only affected by the $\alpha$-cluster probability at
the surface \cite{lyu18}.

\section{Results}
\subsection{Numerical inputs}
In this study, we fix the following kinematical conditions for the
${}^{12}$Be($p$,$p\alpha$)${}^{8}$He reaction in the laboratory frame. The
kinetic energy for the incident and emitted protons are set to be 250 MeV and
180 MeV, respectively. The emission angle of the outgoing proton is set to be
$(\theta_{1}^{\rm{L}},\phi_{1}^{\rm{L}})$= $(60^\circ,0^{\circ})$. To satisfy
the recoilless condition for the $^{8}$He residue, the angle
$\theta_{2}^{\rm{L}}$ of the emitted $\alpha$-cluster varies around
$51^{\circ}$, and the angle $\phi_{2}^{\rm{L}}$ is set as $180^{\circ}$.
The relativistic treatment is adopted in all the reaction kinematics in
this calculation as well as the kinematics of the $p$-$\alpha$ binary
collision. Recently, the importance of the dynamical relativistic corrections
to the Coulomb and nuclear interactions has been revealed for the breakup
reactions \cite{Bertulani05,Ogata09,Ogata10,Long11}. To see the effect of the
dynamical relativistic corrections in the $(p,p\alpha)$ knockout reactions
will be interesting, but it is beyond the scope of current study.

For the Hamiltonian of $^{12}$Be in structural calculation, we adopt the MV1
interaction \cite{ando80} of the central force, which includes finite-range
two-body term and zero-range three-body terms. The two-body spin-orbit term is
adopted from the G3RS interaction \cite{Yamaguchi79}. The parameters in these
interactions and the width $b$ of the Gaussian wave packet adopted here are
those used in Ref.~\cite{enyo03}, where the energy spectra of low lying states
in $^{11}$Be and $^{12}$Be nuclei are well reproduced by the AMD calculations
\cite{enyo02,enyo03}.

\subsection{Energy spectrum of the $^{12}$Be nucleus}
\label{subsec:energy}
\begin{figure}[htb]
  \centering
  \includegraphics[width=\figwidth\textwidth]{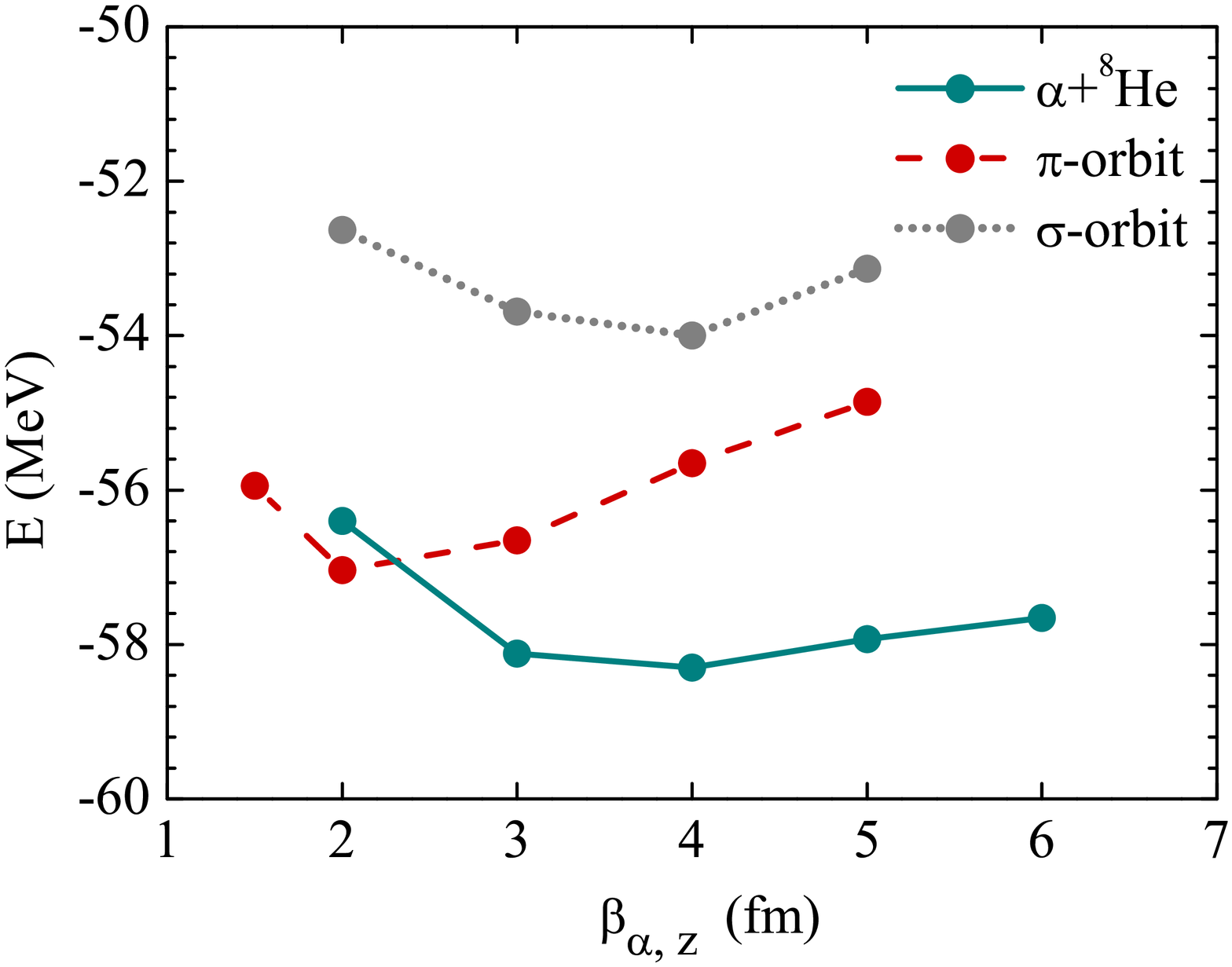}
  \caption{Energy curves versus the variation of parameter $\beta_{\alpha,z}$
    in the $\pi\text{-orbit}$ (dashed), $\sigma\text{-orbit}$ (dotted) and
    $\alpha+{}^{8}\textrm{He}$ (solid) configurations of $^{12}$Be.}
  \label{fig:12be-varbeta}
\end{figure}

We calculate the energy and the wave function of $^{12}$Be by diagonalizing
the Hamiltonian with respect to the basis states formulated in Section
\ref{sec:thsr}. First, we discuss the properties of the bases in each cluster
configuration, that is $\pi$-orbit, $\sigma$-orbit, or
$\alpha+{}^{8}\textrm{He}$. In Fig.~\ref{fig:12be-varbeta}, the energies are
plotted as functions of the parameter $\beta_{\alpha,z}$, which specifies the
spatial extent of cluster motion. One sees clear dependence of the energy on
$\beta_{\alpha,z}$ for all the three configurations. In the $\pi$-orbit
(dashed curve) configuration, the energy minimum locates at about 2 fm, which
corresponds to a very compact $\alpha$-clustering structure due to the
external bounding from valance neutrons in $\pi$-states. In $\sigma$-orbit
(solid curve) and $\alpha+{}^{8}\textrm{He}$ (dotted curve) configurations,
the energy minima locate at much larger $\beta_{\alpha,z}=4$ fm, which
corresponds to very large spatial distribution of $\alpha$-clusters. In
addition, the energies described by the $\pi$-orbit and
$\alpha+{}^{8}\textrm{He}$ configurations are found to be comparable with each
other, which indicates that the breaking of the neutron-magic number $N=8$
could occur through a strong state mixing between these two configurations in
the ground state of $^{12}$Be. The bases in the $\sigma$-orbit configuration
are energetically unfavored compared with the other two configurations, and
could give small contribution to the ground state of $^{12}$Be.

In numerical calculations, we prepare the THSR bases with various parameters
$\beta_{\alpha,z}$ in the $\pi$-orbit, $\sigma$-orbit and
$\alpha+{}^{8}\textrm{He}$ configurations. After the diagonalization of the
Hamiltonian matrix, it is found that the total wave function of the ground
state of $^{12}$Be is efficiently described only by bases of the $\pi$-orbit
and $\alpha+{}^{8}\textrm{He}$ configurations. However, the $\sigma$-orbit
configuration gives negligible contribution because the $\sigma$-orbit bases
have large overlap with the corresponding $\alpha+{}^{8}\textrm{He}$ bases and
its contribution to the ground state can be effectively taken into account by
the $\alpha+{}^{8}\textrm{He}$ configuration in the THSR framework. Therefore,
for the final result of $^{12}$Be, we omit the $\sigma$-orbit bases and adopt
6 bases of the $\alpha+{}^{8}\textrm{He}$ and $\pi$-orbit configurations by
choosing $\beta_{\alpha,z}=2,3,4$ fm for each configuration in the
superposition given by Eq.~(\ref{eq:12be-total}).

\begin{table}[htbp]
  \begin{center}
    \caption{\label{table:diag} The energy ($E$ in MeV) of the ground state
   and excitation energies ($\Delta E$ in MeV) of the $0^+_2$ and $2^+_1$
   states in $^{12}$Be calculated with the superposition of the six bases from
   $\pi$-orbit and $\alpha+{}^{8}\textrm{He}$ configurations. The MV1
   potential and G3RS potential are used for interactions in the central and
   spin-orbit channels, respectively. Parameters of these potentials are
   adopted from Ref.~\cite{enyo03}. ``THSR" denotes the results calculated in
   this study using the THSR wave function. ``AMD'' denotes the results from
   the AMD investigations using the same interaction as in Ref.~\cite{enyo03}.
   ``THSR (weakened $V_{0}^{ls}$)'' denotes the results calculated using the
   same THSR bases as in Eq.~(\ref{eq:12be-total}) but with weakened
   spin-orbit coupling strength $V_{0}^{ls}$=3000 MeV. ``exp.'' denotes
   corresponding experimental values.}
      \begin{tabular}{cccc}
      \hline
      \hline
      $^{12}$Be                 & $E$($0_{1}^{+}$) &$\Delta E$($0_{2}^{+}$)
      &$\Delta E$($2_{1}^{+}$)\\
      \hline
      THSR    & $-$59.5 & 4.1 & 2.2 \\
      AMD     & $-$61.9 & 3.7 & 2.1 \\
      exp.    & $-$68.6 & 2.3 & 2.1 \\
      THSR (weakened $V_{0}^{ls}$) & $-$58.0 & 4.6 & 3.0 \\
      \hline
      \hline
    \end{tabular}
  \end{center}
\end{table}

The calculated energy spectrum for low lying states, $0^+_1$, $0^+_2$, and
$2^+_1$, of $^{12}$Be are listed in Table \ref{table:diag}. For the ground
state, we obtain the binding energy of $-$59.5 MeV, which is somewhat higher
than the experimental value but acceptable as our main purpose is to describe
correctly the cluster wave functions and to reproduce the energy spectrum, not
to precisely reproduce the total energy. Especially, the excitation energy of
the $2_{1}^{+}$ state is very sensitive to the moment of inertia determined by
the distribution of the $\alpha$-clusters. In this work, we reproduce well the
energy gap for the ground band as $\Delta E(2_{1}^{+})$=2.2 MeV, which is
consistent with the experimental value 2.1 MeV. As a comparison, we also show
the results calculated using the same THSR bases as in
Eq.~(\ref{eq:12be-total}) but with a weakened spin-orbit coupling strength
$V_{0}^{ls}$=3000 MeV (the default value is 3700 MeV in Ref.~\cite{enyo03}).
The calculation with the weakened spin-orbit strength shows the higher
excitation energy of $\Delta E(2_{1}^{+})$=3.0 MeV than the experimental value
and may indicate a weaker $\alpha$-clustering in the ground band. The
excitation energy calculated for the $0_{2}^{+}$ state is 4.1 MeV with the
default parameter in this study, which is consistent with the corresponding
AMD result of 3.7 MeV but still higher than the experimental value.

\subsection{Mixture of configurations in the $^{12}$Be nucleus}
\label{subsec:mix}
The mixture of the $\pi$-orbit and $\alpha+{}^{8}\textrm{He}$ configurations
is explicitly treated as shown in Eq.~(\ref{eq:12be-total}). In order to
discuss the contribution from the cluster configurations to the total wave
function of $^{12}$Be, we define a probability to find each configuration in
the total wave function by
\begin{equation}\label{eq:overlap}
  P_{m}=\frac
  {|\braket{\Psi(^{12}\text{Be})|\Phi'_m}|^{2}}
  {
    \braket{\Psi(^{12}\text{Be})|\Psi(^{12}\text{Be})}
    \braket{\Phi'_m|\Phi'_m}
  },
\end{equation}
where $\Psi(^{12}\text{Be})$ is the total wave function of $^{12}$Be in
Eq.~(\ref{eq:12be-total}) and $\Phi'$ is defined by 
\begin{equation}
    \Phi'_{m}= \sum_{j=1,2,3} c_{m,j}
    \hat{P}^{0}_{00}\hat{P}_{\text{c.o.m}}\Phi_{m}(\beta_{\alpha,z;j})
\end{equation}  
with $\beta_{\alpha,z;\{j=1,2,3\}}=\{2,3,4$ fm\} and the label $m$ denotes
$\pi$-orbit or $\alpha+{}^{8}\textrm{He}$ within the two configurations. Here
the coefficients $c_{i}$ are fixed to be the values determined by the full
diagonalization for 6 bases.

\begin{figure}[htb]
  \centering
  \includegraphics[width=\figwidth\textwidth]{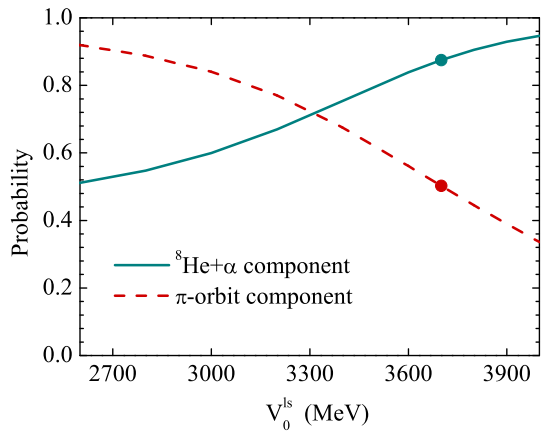}
  \caption{The probability to find the $\alpha$+$^{8}$He (solid curve) and
  $\pi$-orbit (dashed curve) components in the total wave function for the
  ground state of $^{12}$Be. The horizontal axis is the spin-orbit coupling
  strength $V_{0}^{ls}$. The solid dots are the values correspond to the
  default choice of $V_{0}^{ls}$.}
  \label{fig:overlap}
\end{figure}

In Fig.~\ref{fig:overlap}, we show the probability of each component in the
total wave function of $^{12}$Be. The probabilities of the $\pi$-orbit (dashed
curve) and $\alpha$+$^{8}$He (solid curve) components are plotted as function
of the spin-orbit coupling strength $V_{0}^{ls}$. A strong dependence on
$V_{0}^{ls}$ is observed for the mixing ratio between the two configurations.
As the spin-orbit coupling strength increases, the $\alpha$+$^{8}$He component
increases because the $\alpha$+$^{8}$He configuration comes to the energy
relatively lower than the $\pi$-orbit configuration as shown in
Fig.~\ref{fig:12be-varbeta}. As a consequence, the dominant component is
changed from the $\pi$-orbit to the $\alpha$+$^{8}$He configuration, which
simulates the gradual transition from the normal state to the intruder state
in the ground state wave function. With the default choice of
$V_{0}^{ls}$=3700 MeV, the ground state of $^{12}$Be contains a 90\%
$\alpha$+$^{8}$He component and have a largely developed $\alpha$-clustering.
With a weakened spin-orbit coupling strength $V_{0}^{ls}$=3000 MeV, the
$\alpha$+$^{8}$He component reduces significantly to about 60\% corresponding
to the modest $\alpha$-clustering. It should be noted that the $\pi$-orbit and
$\alpha$+$^{8}$He configurations are not orthogonal to each other and the
ground state also has 50\% and 80\% $\pi$-orbit probabilities for the default
and weakened $V_{0}^{ls}$ cases, respectively. Considering that the
$V_{0}^{ls}$ and other parameters in the $NN$ interactions are model dependent
in different microscopic calculations, the ambiguities are inevitable for the
mixing ratios between clustering configurations. Hence, the experimental
observables that are directly related to these mixing ratios are essential for
the validation of the predictions from the nuclear theories.

\subsection{The $\alpha$-cluster wave function}
\begin{figure}[htb]
  \centering
  \includegraphics[width=\figwidth\textwidth]{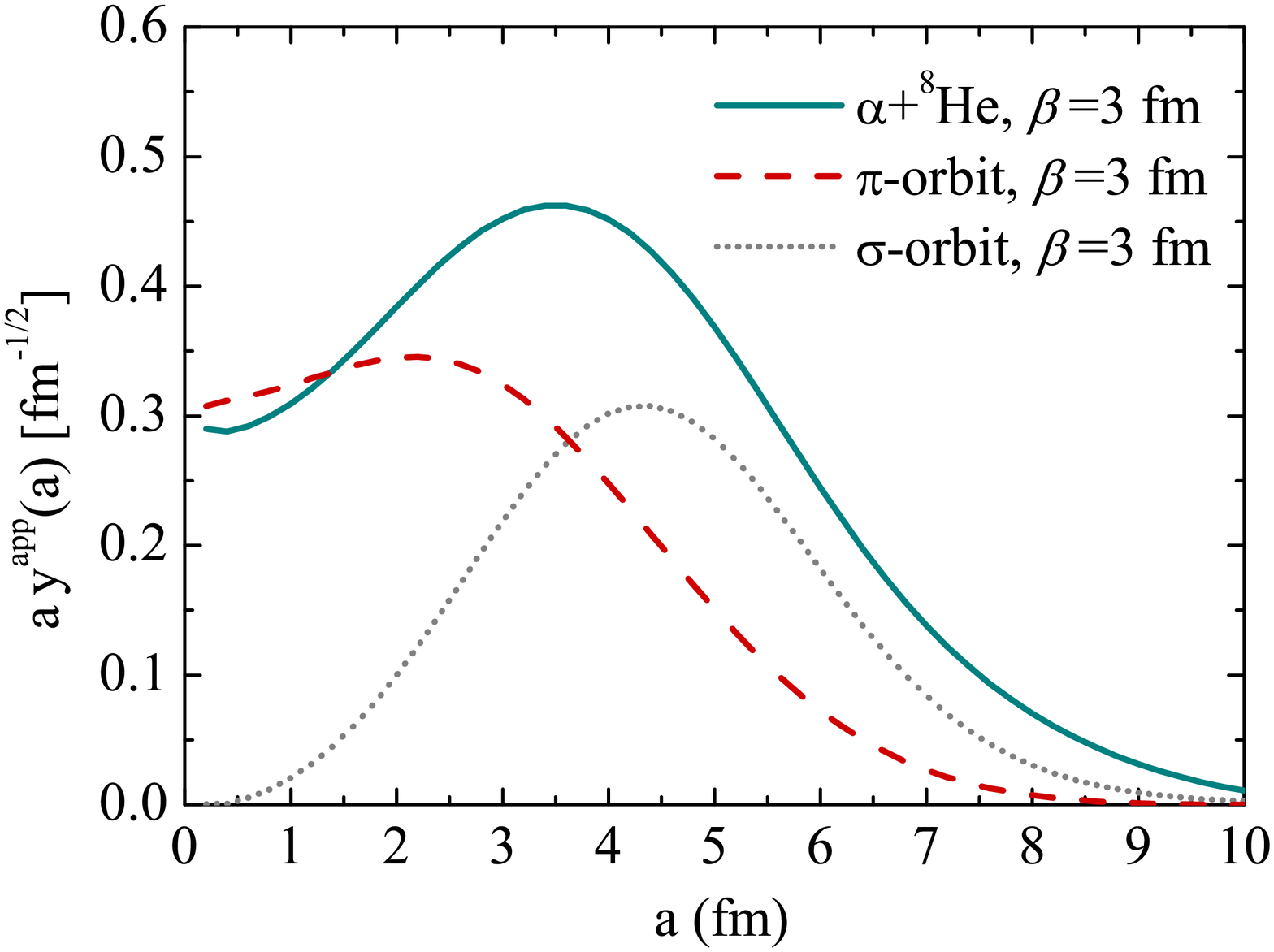}
  \caption{\label{fig:rwa-chan}Comparison of approximated RWAs for the THSR
  bases in the $\alpha+{}^{8}\textrm{He}$ (solid curve), $\pi$-orbit (dashed
  curve), and $\sigma$-orbit (dotted curve) configurations of $^{12}$Be.
  Formulations of each basis are introduced in the text and the parameter
  $\beta_{\alpha,z}$ is set to be 3 fm.}
\end{figure}

The $\alpha$-cluster wave function of $^{12}$Be can be obtained with the
approximated RWA as described in Sec \ref{subsec:rwa}. In
Fig.~\ref{fig:rwa-chan}, we compare the approximated RWAs for the THSR bases
in the $\alpha+{}^{8}\textrm{He}$ (solid curve), $\pi$-orbit (dashed curve),
and $\sigma$-orbit (dotted curve) configurations. It is clearly shown that the
$\alpha+{}^{8}\textrm{He}$ configuration shows much larger amplitude at the
surface region because it describes the enhanced $\alpha$-clustering comparing
to in the $\pi$-orbit configurations. We note again that the cross sections of
the $\alpha$-knockout reaction are sensitive to the $\alpha$ amplitudes in the
surface region but are not affected by the amplitudes in the inner region.

\begin{figure}[htb]
  \centering
  \includegraphics[width=\figwidth\textwidth]{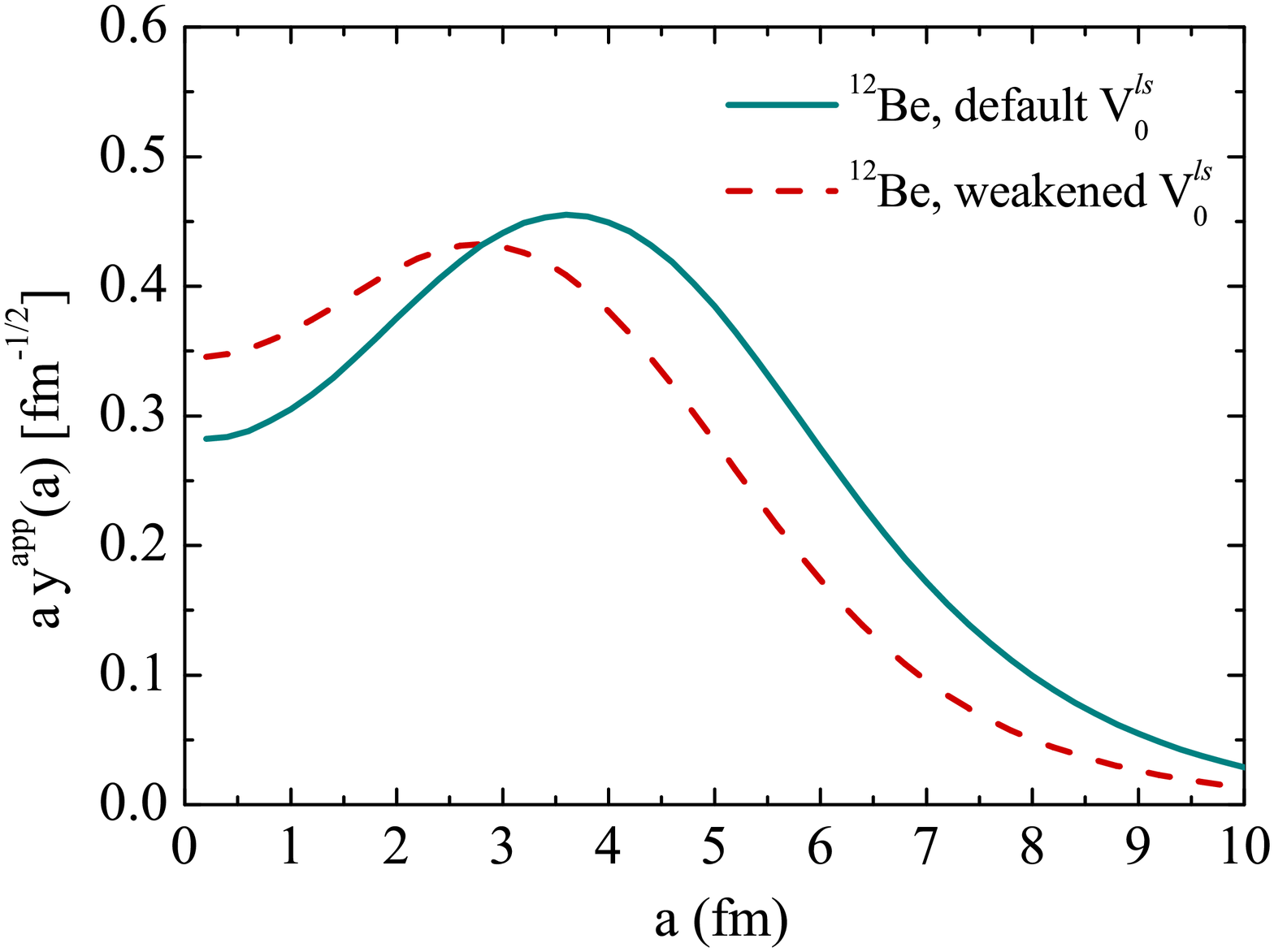}
  \caption{\label{fig:rwa}The approximated RWAs for the $^{12}$Be target with
  the default interaction (solid curve) and the weakened spin-orbit coupling
  strength (dashed curve).}
\end{figure}

In Fig.~\ref{fig:rwa}, we compare the approximated RWAs for the $^{12}$Be
target with the default interaction (solid curve) and that with the weakened
spin-orbit coupling strength (dashed curve). In the surface region, a
significant difference is observed for the $\alpha$ amplitudes between the
curves. The calculation with the default interaction gives larger surface
amplitude than with the weakened spin-orbit coupling strength because of the
larger $\alpha$+$^{8}$He component. As shown in Sec.~\ref{subsec:tdx}, this
difference in the $\alpha$ amplitudes in the surface region can be examined by
the TDX observables in the $\alpha$-knockout reaction.

\subsection{The triple differential cross sections}
\label{subsec:tdx}
\begin{figure}[htb]
  \centering
  \includegraphics[width=\figwidth\textwidth]{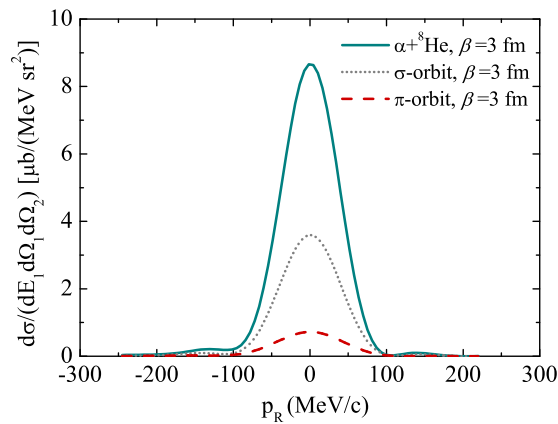}
  \caption{Comparison of TDXs calculated using one THSR basis in each of the
  $\alpha+{}^{8}\textrm{He}$, $\sigma$-orbit and $\pi$-orbit configurations.
  Parameters $\beta_{\alpha,z}$ in these bases are set to be 3 fm. The same
  kinematics is adopted as in Fig.~\ref{fig:tdx}.}
  \label{fig:tdx-channel}
\end{figure}

In Fig.~\ref{fig:tdx-channel}, the TDXs are compared for the THSR bases in the
$\alpha+{}^{8}\textrm{He}$ (solid curve), $\sigma$-orbit (dotted curve), and
$\pi$-orbit (dashed curve) configurations with $\beta_{\alpha,z}=3$ fm. A
prominent TDX is obtained for the solid curve, which is a logical outcome of
the strong $\alpha$-clustering in the $\alpha+{}^{8}\textrm{He}$
configuration. On the other hand, the dashed curve has a significantly lower
peak height, which is consistent to the weak $\alpha$-clustering in the
$\pi$-orbit configuration. The huge difference in the magnitude with a factor
of 10 between these two configurations indicates that the ($p,p\alpha$)
reactions could be used as a sensitive tool to differentiate the mixing of the
strong and week clustering components. For the $\sigma$-orbit configuration,
we note that the dashed curve in Fig.~\ref{fig:tdx-channel} shows the TDX with
about half magnitude of the solid curve, as expected from the RWA shown in
Fig.~\ref{fig:rwa-chan}, where a ratio of about 0.5 is obtained for the
squared values between the $\sigma$-orbit and the $\alpha+{}^{8}\textrm{He}$
configurations.

\begin{figure}[htb]
  \centering
  \includegraphics[width=\figwidth\textwidth]{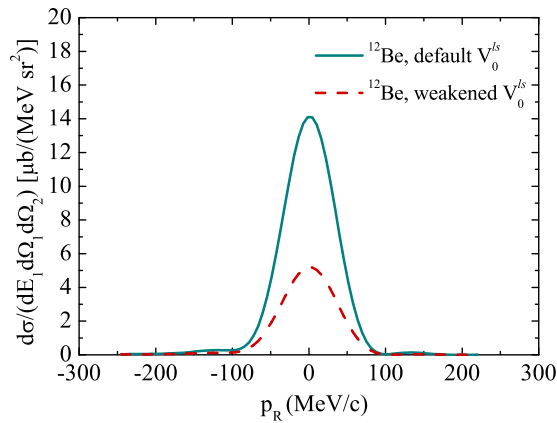}
  \caption{The TDXs of the ${}^{12}$Be($p$,$p\alpha$)${}^{8}$He reaction at
  250 MeV, predicted by calculations using the default and the weakened
  spin-orbit coupling strength. Kinetic energy of particle 1 is fixed at 180
  MeV and its emission angle is set to $(\theta_1^{\rm{L}},
  \phi_1^{\rm{L}})=(60^\circ, 0^\circ)$. $\phi_2^{\rm{L}}$ is fixed at
  $180^\circ$ and $\theta_2^{\rm{L}}$ is varied around $51^\circ$. $P_R$ is
  the recoiled momentum.}
  \label{fig:tdx}
\end{figure}

In Fig.~\ref{fig:tdx}, the theoretical predictions of the TDXs are shown. When
default $V_{0}^{ls}$ is adopted in the $NN$ interaction, as shown by the solid
curve in Fig.~\ref{fig:tdx}, the TDXs are found to be analogous to the values
of the $\alpha+{}^{8}\textrm{He}$ configuration in Fig.~\ref{fig:tdx-channel}.
In this case, the neutron magic number $N=8$ apparently breaks because of the
intruder occupation induced by the $\alpha$-cluster formation. The dashed
curve in Fig.~\ref{fig:tdx} corresponds to the weakened $V_{0}^{ls}=3000$ MeV,
where the intermediate strength of $\alpha$-cluster formation is suggested by
the probability calculation in Fig.~\ref{fig:overlap}, and we expect weaker
breaking of $N=8$ than in the default case. The ratio of about 2 is observed
for the TDXs between the default and weakened curves at the zero momentum. We
stress again that the TDX curves are sensitive to the $\alpha$-clustering in
the wave function. In particular, this difference is much larger than in the
RWA curves in Fig.~\ref{fig:rwa}. 

In both Figs.~\ref{fig:tdx-channel} and \ref{fig:tdx}, the high sensitivities
of the ($p,p\alpha$) reaction are established for clarifying the strong and
the week $\alpha$-clustering. We credit this superiority to the peripheral
property of the ($p,p\alpha$) reaction \cite{yoshida16,yoshida18}, which
allows probing of the $\alpha$-clusters only in the surface region where the
probability of cluster formation is the largest. Hence, by comparing the
experimental values of TDX with the theoretical predictions in
Figs.~\ref{fig:tdx-channel} and \ref{fig:tdx}, we can validate the breaking of
$N=8$ by cluster formation. Furthermore, differentiation between the strong
and the weak $\alpha$-clustering in the ground state of $^{12}$Be will be
feasible.

Recently, there have been several works regarding the use of eikonal
scattering waves in the DWIA framework for the $(p,2p)$ reactions
\cite{Aumann13, Atar18,Liu19}. It will be interesting to evaluate the
efficiency of eikonal approximation in the current $(p,p\alpha)$ case, which
is expected to be discussed in our future work.

\section{Summary}
We have provided the direct probing for the $\alpha$-clustering structures in
the ground state of $^{12}$Be nucleus through the $(p,p\alpha)$ reaction at
250 MeV. The target and residual nuclei are described by the new framework of
a nonlocalized cluster model with valence neutrons, and the reaction process
is treated by the DWIA framework. The rich phenomena in the low-lying states
of $^{12}$Be target, such as the coexistence of binary cluster and MO
configurations, are described by the superposition of bases extending the THSR
wave functions. The low-lying energy spectrum and the probabilities of strong
and weak clustering components in the ground state of $^{12}$Be are obtained
by the structural calculations using the newly formulated wave function. It is
found that the magic number $N=8$ breaking occurs because of the strong
clustering in the ground state of $^{12}$Be. The huge difference in the
magnitude of TDXs at the zero momentum between the $\pi$-orbit and the
$\alpha+{}^{8}\textrm{He}$ configurations shows that the TDX is a good measure
for the breaking of $N=8$. In addition, the TDX is found to be highly
sensitive to the strong and weak cluster formations, which allows quantitative
discussions for the corresponding mixing ratio in the ground state of
$^{12}$Be. This study provides a feasible approach to probe directly the
exotic clustering features in the ground state of $^{12}$Be. Furthermore, the
new THSR wave function formulated in this work provides new option for the
study of neutron rich nuclei near the drip-line.

\begin{acknowledgments}
The authors thank K.~Minomo, Y.~Neoh, Y.~Chazono, and N.~Itagaki for valuable
discussions. The computation was carried out with the computer facilities at
the Research Center for Nuclear Physics, Osaka University. This work was
supported in part by Grants-in-Aid of the Japan Society for the Promotion of
Science (Grants No. JP16K05352, No. JP15J01392, and No. JP18K03617). 
\end{acknowledgments}

\appendix*
\section{Formulations of cluster and molecular orbit states}
We prepare the cluster and the molecular orbit states by the new extension of
the THSR formulations used in our previous work \cite{lyu15, lyu16}.

\subsection{The $\alpha$-cluster and the $^{8}$He-cluster}
In order to simplify the discussion, we define the Gaussian wave packet in
real space for nucleons as
\begin{equation}
  g(\bm{r},\bm{R})= 
    \left(\frac{1}{\pi b^{2}}\right)^{3/4}
    \exp\left\{-\frac{(\bm{r}-\bm{R})^2}{2b^{2}}\right\},
\end{equation}
and the $s$-states of nucleons are written as the product of the spatial wave
packet and the spin-isospin term
\begin{equation}\label{eq:s-state}
  \phi^{s}_{\tau s}(\bm{r},\bm{R})=g(\bm{r},\bm{R})\ket{\tau,s}.
\end{equation}
The $\alpha$-clusters are described by the antisymmetrization of four
$s$-states with spin-isospin saturation, as
\begin{equation}
  \Psi_{\alpha}(\bm{R})=\frac{1}{\sqrt{4!}}
    \mathcal{A}\{\phi_{1}\phi_{2}\phi_{3}\phi_{4}\},
\end{equation}
where
\begin{equation}
  \begin{aligned}
    \phi_{1}=\phi^{s}_{p\uparrow}(\bm{r},\bm{R}),\quad
     & \phi_{2}=\phi^{s}_{p\downarrow}(\bm{r},\bm{R}), \\
    \phi_{3}=\phi^{s}_{n\uparrow}(\bm{r},\bm{R}),\quad
     & \phi_{4}=\phi^{s}_{n\downarrow}(\bm{r},\bm{R}).
  \end{aligned}
\end{equation}

The $^{8}$He-cluster wave function is written as the Slater determinant of
eight single nucleon states, including four $s$-states in $\alpha$-cluster and
four surrounding $p$-states, as
\begin{equation}\label{eq:8he-single-states}
  \begin{aligned}
    \ket{\Phi(^{8}\textrm{He},\bm{R})}
     & =\frac{1}{\sqrt{8!}}
    \ket{\mathcal{A}\{
      \phi_{5}\phi_{6}\phi_{7}\phi_{8}
      \phi_{9}\phi_{10}\phi_{11}\phi_{12} \}
    }                       \\
     & =\frac{1}{\sqrt{8!}}
    \ket{\mathcal{A}\{
      \phi_{\alpha}(\bm{R})\phi_{9}\phi_{10}\phi_{11}\phi_{12} \}
    },
  \end{aligned}
\end{equation}
where $\phi_{5\mbox{--} 8}$ denote the $s$-states and $\phi_{9\mbox{--} 12}$
denote the $p$-states. The $\phi_{9,10}$ correspond to the $1P_{3/2,\pm 3/2}$
states with the ring-type distribution on the horizontal plane, and they are
simulated by the integration
\begin{equation}
  \label{eq:p-orbit}
  \begin{aligned}
    \phi_{9,10}(\bm{r},\bm{R})=
     & \int d\bm{R}'
    \mathcal{G}(\bm{R}',\bm{\beta}_{p})
    e^{\pm i\phi_{\mathbf{R}'}}
    g(\bm{r},\bm{R}+\bm{R}')         \\
     & \phantom{\int d\bm{R}'}\times
    \ket{\tau,\sigma=\pm 1/2}.
  \end{aligned}
\end{equation}
In the limit of $\bm{\beta}_{p} \to \bm{0}$ fm, the states $\phi_{9,10}$
converge to the $1P_{3/2,\pm 3/2}$ states of the harmonic oscillators. For the
other two $p$-states in the $^{8}$He cluster, we project the desired
$1P_{3/2,\pm 1/2}$ states from the vertical rotation of states $\phi_{9,10}$,
as
\begin{equation}
  \label{eq:p-orbit-vertical}
  \begin{aligned}
    \phi_{11}(\bm{r},\bm{R})
     & =\hat{R}(\Omega)\phi_{9}(\bm{r},\bm{R}), \\
    \phi_{12}(\bm{r},\bm{R})
     & =\hat{R}(\Omega)\phi_{10}(\bm{r},\bm{R}),
  \end{aligned}
\end{equation}
where the Euler angle $\Omega$ is $\{0,\pi/2,0\}$. Because of the total
antisymmetrization between the four neutron states $\phi_{9\mbox{--} 12}$ in
the $1P_{3/2}$ orbits, only the $1P_{3/2,\pm 1/2}$ components of the rotated
states $\phi_{11,12}$ contribute to the total cluster wave function of
$^{8}$He.

\subsection{The $\pi$-orbit states}
The $\pi$-orbits are written as \cite{lyu15}
\begin{equation}
  \label{eq:pi-orbit}
  \begin{aligned}
    \phi^{\pi}_{\tau s}
    (\bm{r})=\int d\bm{R}'
    \mathcal{G}(\bm{R}',\bm{\beta}_{\pi})
    e^{\pm i\phi_{\bm{R}'}}
    g(\bm{r}_{i},\bm{R}')
    \ket{\tau, s=\pm 1/2},
  \end{aligned}
\end{equation}
and
\begin{equation}
  \label{eq:pi-orbit2}
  \begin{aligned}
    \phi^{\pi*}_{\tau s}
    (\bm{r})=\int d\bm{R}'
    \mathcal{G}(\bm{R}',\bm{\beta}_{\pi}')
    e^{\pm i\phi_{\bm{R}'}}
    g(\bm{r}_{i},\bm{R}')
    \ket{\tau, s=\mp 1/2},
  \end{aligned}
\end{equation}
where the superscripts $\pi$ and $\pi*$ denote the $\pi$-orbits with parallel
and antiparallel spin-isospin coupling, respectively. The states
$\phi^{\pi}_{9\mbox{--} 12}$ of four neutrons occupying $\pi$-orbits in
Eqs.~(\ref{eq:12Be-pi}) and (\ref{eq:12Be-sigma}) are defined as
\begin{equation}
  \begin{aligned}
    \phi^{\pi}_{9}=\phi^{\pi}_{n\uparrow}(\bm{r}),\quad
     & \phi^{\pi}_{10}=\phi^{\pi}_{n\downarrow}(\bm{r}),    \\
    \phi^{\pi*}_{11}=\phi^{\pi*}_{n\uparrow}(\bm{r}),\quad
     & \phi^{\pi*}_{12}=\phi^{\pi*}_{n\downarrow}(\bm{r}).
  \end{aligned}
\end{equation}

\subsection{The $\sigma$-orbit states}
The $\sigma$-orbits in $^{12}$Be nucleus are formulated with respect to the
$\alpha$-clusters, as
\begin{equation}
  \begin{aligned}
    \label{eq:sigma-single}
    \phi^{\sigma}_{\tau s}(\bm{r}, \bm{R})=
     & \int d\bm{R}' \mathcal{G}(\bm{R}',\bm{\beta}_{\sigma})
    \mathcal{F}(\bm{R})\mathcal{F}(\bm{R}')                   \\
     & \quad \times g(\bm{r},\bm{R}+\bm{R}')\ket{\tau,s}.
  \end{aligned}
\end{equation}
where $\pm\bm{R}$ are the generate coordinates of two $\alpha$-clusters in
Eq.~(\ref{eq:12Be-sigma}). The factor functions $\mathcal{F}$ are defined by
\begin{equation}
  \label{eq:f-function}
  \mathcal{F}(\bm{R})=
  \begin{cases}
    +1 \quad (R_{z}>0) \\
    -1 \quad (R_{z}<0) \\
  \end{cases}.
\end{equation}
In Eq.~(\ref{eq:12Be-sigma}), there is integration over the $\alpha$-cluster
generate coordinate $\bm{R}$, as
\begin{equation}
  \begin{aligned}
    \label{eq:sigma-integrated}
     & \int d\bm{R} \mathcal{G}(\bm{R},\bm{\beta})
    \phi^{\sigma}_{\tau s}(\bm{r},\bm{R})          \\
    =
     & \int d\bm{R} d\bm{R}'
    \mathcal{G}(\bm{R},\bm{\beta})
    \mathcal{G}(\bm{R}',\bm{\beta}_{\sigma})
    \mathcal{F}(\bm{R})\mathcal{F}(\bm{R}')        \\
     & \quad\times g(\bm{r},\bm{R}'+\bm{R})
    \ket{\tau,s},
  \end{aligned}
\end{equation}
which numerically describes the single nucleon state in $\sigma$-orbit
configuration, as shown by corresponding density distribution in
Fig.~\ref{fig:dens} (b). The states $\phi_{11,12}^{\sigma}$ of two neutrons
occupying the $\sigma$ orbits in Eq.~(\ref{eq:12Be-sigma}) are defined as
\begin{equation}
  \phi^{\sigma}_{11}=\phi^{\sigma}_{n\uparrow}(\bm{r},\bm{R}),\quad
  \phi^{\sigma}_{12}=\phi^{\sigma}_{n\downarrow}(\bm{r},\bm{R}).
\end{equation}

\subsection{Parameters of the THSR bases}
We list in Table \ref{table:para} the parameters of the THSR basis states used
in the numerical calculation. In the table, $\beta_{\alpha}$s are parameters
for the cluster motion in each configuration as shown in
Eqs.~(\ref{eq:12Be-2cluster}), (\ref{eq:12Be-pi}) and (\ref{eq:12Be-sigma}).
For the $\alpha+{}^{8}\textrm{He}$ configuration, $\beta_{n}$s denote
parameters $\beta_{p}$s for the neutrons occupying the $p$-orbits in the
$^{8}$He cluster, as shown in Eqs.~(\ref{eq:p-orbit}) and
(\ref{eq:p-orbit-vertical}). For the $\pi$-orbit and $\sigma$-orbit
configurations, $\beta_{n}$s denote parameters $\beta_{\pi}$s and
$\beta_{\sigma}$s respectively for neutrons occupying the MO orbits, as shown
in Eqs.~(\ref{eq:pi-orbit}), (\ref{eq:pi-orbit2}) and (\ref{eq:sigma-single}).
The superscripts $9\mbox{--} 12$ denote parameters for corresponding single
neutron states $\phi_{9\mbox{--} 12}$ in each configuration.

\begin{table}[htbp]
  \begin{center}
    \caption{\label{table:para} Parameters of the THSR basis states in the
    $\alpha+{}^{8}\textrm{He}$, $\pi$-orbit, and $\sigma$-orbit
    configurations. Detailed explanations of the parameters are given in the
    main text. All units are in fm.}
      \begin{tabular}{ccccccc}
      \hline
      \hline
      basis &$\beta_{\alpha,xy}$      &$\beta_{\alpha,z}$
      &$\beta_{n,xy}^{9,10}$ &~$\beta_{n,z}^{9,10}$ &~$\beta_{n,xy}^{11,12}$
      &~$\beta_{n,z}^{11,12}$\\
      
      \hline
      $\Phi_{\alpha+{}^{8}\text{He}}$ & 0.1 & ~2.0, 3.0, 4.0~ & 0.1 & 0.4 &
      0.1 & 0.4 \\
      $\Phi_{\pi\text{-orbit}}$ & 0.1 & ~2.0, 3.0, 4.0~ & 1.5 & 3.0 & 2.5 &
      3.0 \\
      $\Phi_{\sigma\text{-orbit}}$ & 0.1 & ~2.0, 3.0, 4.0~ & 1.5 & 3.0 & 0.1 &
      2.0 \\
      \hline
      \hline
    \end{tabular}
  \end{center}
\end{table}


\end{document}